\documentstyle[preprint,aps,version2,epsfig]{revtex}
\draft
\def\Journal#1#2#3#4{{#1} {\bf#2}, {#4} {(#3)}}

\def\NP{{ Nucl. Phys.} }

\def\PL{{ Phys. Lett.}}

\def\PRP{{ Phys. Rep.}}
\def\PRL{ Phys. Rev. Lett.}
\def\PR{{ Phys. Rev.}}

\def\ZP{{Z. Phys.}}

\def\EPJ{{Eur. Phys. J.}}

\def\MPL{{Mod. Phys. Lett.}}

\def\IJMP{Int. J. Mod. Phys.}

\def\ra{\rightarrow}

\def\be{\begin{equation}}
\def\ee{\end{equation}}
\def\bea{\begin{eqnarray}}
\def\eea{\end{eqnarray}}

\def\ua{\uparrow}
\def\da{\downarrow}
\def\qbar{{\bar q}}
\def\ubar{{\bar u}}
\def\dbar{{\bar d}}
\def\sbar{{\bar s}}

\def\uup{{u^\uparrow}}
\def\udn{{u^\downarrow}}
\def\dup{{d^\uparrow}}
\def\ddn{{d^\downarrow}}

\def\ubarup{{\bar u}^\uparrow}
\def\ubardn{{\bar u}^\downarrow}
\def\dbarup{{\bar d}^\uparrow}
\def\dbardn{{\bar d}^\downarrow}

\def\NP{{ Nucl. Phys.}}
\def\ANP{{Adv. Nucl. Phys.}}

\begin{document}
\begin{titlepage} 

\bigskip
\begin{center}
{\large\bf 
The flavour asymmetry of polarized anti-quarks
\\ in the nucleon}

\author{Fu-Guang Cao\thanks{E-mail address: f.g.cao@massey.ac.nz.}
 and A. I. Signal\thanks{E-mail address: a.i.signal@massey.ac.nz.}}
\begin{instit}
Institute of Fundamental Sciences \\ Massey University \\
Private Bag 11 222,  Palmerston North \\
New Zealand
\end{instit}
\end{center}

\begin{abstract}
We present a study of the flavour asymmetry of polarized anti-quarks
in the nucleon using the meson cloud model.
We include contributions both from the vector mesons and the interference terms
of pseudoscalar and vector mesons.
Employing the bag model, we first give the polarized valence quark distribution
of the $\rho$ meson and the interference distributions.
Our calculations show that the interference effect mildly increases the prediction
for $\Delta \dbar(x)-\Delta \ubar(x)$ at intermediate $x$ region.
We also discuss the contribution of `Pauli blocking' to the asymmetry.


\end{abstract}
\end{titlepage}
\setcounter{footnote}{0}
\section{Introduction}

The possible breaking of parton model symmetries by the nucleon's quark 
distribution functions has been a topic of great interest since the experimental 
discoveries that the Ellis-Jaffe \cite{EMC} and Gottfried \cite{NMC} sum rules 
are violated. In particular, the flavour asymmetry in the nucleon sea 
($\dbar > \ubar$) has been confirmed by several experiments \cite{NA51,E866}, 
and the $x$-dependence of this asymmetry has been investigated.
This asymmetry can be naturally explained in the meson cloud model,
in which the physical nucleon wave function contains many virtual meson-baryon 
components, and the valence anti-quark in the meson contributes (via a convolution)  
to the anti-quark distributions in the proton sea.
Since the probability of the Fock state $|n\pi^+\rangle$  is larger than that of the
$|\Delta^{++}\pi^-\rangle$ state in the proton wave function, the asymmetry 
$\dbar >\ubar$ emerges naturally in the proton sea. 
There have been many theoretical investigations
(see e.~g. \cite{Thomas83,HHoltmannSS,Speth,Kumano} and references therein)
on this subject.

Recently there has been increasing interest in the question of whether this 
asymmetry extends also to the polarized sea distributions 
{\it i.e.} $\Delta \dbar(x) \neq \Delta\ubar(x)$.
Such a polarized sea asymmetry would make a direct contribution to the Bjorken 
sum rule.
Although well established experimental evidence for a polarized sea asymmetry
is still lacking, some experimental studies have been done \cite{PAsymmetryExp}.
Moreover several parameterizations \cite{PPDFit} for the polarized parton 
distributions arising from fits of the world data from polarized experiments leave 
open the possibility of this asymmetry.
There have also been some theoretical studies on this asymmetry.
In Ref.~\cite{LargeNC1,LargeNC2}, the polarized sea asymmetries are calculated
in the chiral quark-soliton model (using the large-$N_C$ limit). 
Sizable results for $\Delta \dbar(x)-\Delta \ubar(x)$ and 
$\Delta \dbar(x) +\Delta \ubar(x) -2\Delta \sbar(x)$ were found, and it was further 
predicted that the flavour asymmetry of the polarized sea distributions is larger than
that of the unpolarized sea distributions, {\it i.e.}
$\left|\Delta \dbar(x)-\Delta\ubar(x) \right| > \left| \dbar(x)-\ubar(x) \right|$.
Such sizeable asymmetries would make an important contribution (around 20\%) to 
the Bjorken sum rule. 
Fries and Sch\"{a}fer \cite{RFriesS} calculated the non-strange polarized sea asymmetry 
by considering the $\rho$ meson cloud in the meson cloud model. 
Their prediction for  $\Delta \dbar(x) - \Delta \ubar(x)$ is more than one order of 
magnitude smaller than the result from the chiral quark-soliton model.
Boreskov and Kaidalov \cite{Regge} analysed this asymmetry 
by calculating the Regge cut contribution to the imaginary part of
the high-energy photon-nucleon scattering amplitude.
They found that the interference between the amplitudes for the photon coupling 
to a pion or to a rho meson can provide a sizable polarized anti-quark asymmetry
in the small $x$ region.
This asymmetry has also been discussed in the instanton model \cite{Instanton}
and a statistical model \cite{Statistical} for the parton distributions of the nucleon.

In this paper we investigate the flavour asymmetry of the non-strange polarized 
anti-quarks using the meson cloud model. 
We include both the vector meson cloud and the interference terms of the 
pseudoscalar and the vector mesons.
Such interference terms appear naturally in the meson cloud model.
In section II, we derive the formulas in the meson cloud model to calculate
the flavour asymmetry of non-strange polarized anti-quark distributions.
The numerical results are given in section III along with discussion. 
In section IV we discuss how a contribution to the flavour asymmetry can arise from
the Fermion nature of the quarks, often called Pauli blocking.
Section V is a summary.

\section{Flavour asymmetry in the meson cloud model}

In the meson cloud model (MCM) the nucleon can be viewed as a bare nucleon
plus some meson-baryon Fock states which result from the fluctuation
$N \ra M B$.
The wavefunction of the nucleon can be written as \cite{HHoltmannSS},
\bea
|N\rangle_{\rm physical} =  Z |N\rangle_{\rm bare}
+\sum_{MB} \sum_{\lambda \lambda^\prime} 
\int dy \, d^2 {\bf k}_\perp \, \phi^{\lambda \lambda^\prime}_{MB}(y,k_\perp^2)
\, |M^\lambda(y, {\bf k}_\perp); B^{\lambda^\prime}(1-y,-{\bf k}_\perp)
\rangle 
\label{NMCM}
\eea
where $Z$ is the wave function renormalization constant,
$\phi^{\lambda \lambda^\prime}_{MB}(y,k_\perp^2)$ 
is the wave function of the Fock state containing a meson ($M$)
with longitudinal momentum fraction $y$, transverse momentum ${\bf k}_\perp$,
and helicity $\lambda$,
and a baryon ($B$) with momentum fraction $1-y$,
transverse momentum $-{\bf k}_\perp$, and helicity $\lambda^\prime$.
The model assumes that the lifetime of a virtual baryon-meson Fock state is much
larger than the interaction time in the deep inelastic or Drell-Yan
process, thus the quark and anti-quark in the virtual meson-baryon Fock states
can contribute to the parton distributions of the nucleon.
For spin independent parton distributions these non-perturbative contribution 
can usually be expressed as a convolution of fluctuation functions with the 
valance parton distributions in the meson and/or baryon.
For polarised parton distributions in the model it is necessary to 
include all the terms which can lead to the same final state \cite{SchT}. 
This allows the possibility of interference terms between different terms in the 
nucleon wavefunction eq.~(\ref{NMCM}). 
The effect of interference between $N\pi$ and $\Delta\pi$ terms on polarised 
quark distributions was calculated in \cite{CBorosT,SMST}. 
For polarised anti-quark distributions the interference will be between terms 
with different mesons and the same baryon e.g. $N\pi$ and $N\rho$, see 
Fig.~1.

We can write the total meson cloud contribution to the distribution of anti-quarks 
of a given flavour with helicity $\sigma$ as 
\bea
x \delta\qbar^\sigma (x)=\sum_\lambda \int^1_x dy
f^\lambda_{(M_1 M_2)B/N} (y) \frac{x}{y}
\qbar^\sigma_{(M_1 M_2)\lambda}(\frac{x}{y})
\label{xqbar}
\eea
where
\bea
f^\lambda_{(M_1 M_2)B/N} (y)=\sum_{\lambda^\prime}
\int^\infty_0 d k_\perp^2
\phi^{\lambda \lambda^\prime}_{M_1 B}(y, k_\perp^2)
\phi^{*\,\lambda \lambda^\prime}_{M_2 B}(y, k_\perp^2),
\label{ff}
\eea
is the helicity dependent fluctuation function.
The second meson ($M_2$) could be the same as or different from
the first meson ($M_1$). 

For simplicity we denote Eq.~(\ref{xqbar}) as
\bea
x \delta\qbar^\sigma =\sum_\lambda
f^\lambda_{(M_1 M_2)B/N} \otimes \qbar^\sigma_{(M_1 M_2) \lambda}.
\label{xqbar2}
\eea
The two mesons appearing in Eq.~(\ref{xqbar2})
may be both vector mesons ($V$) or one pseudoscalar meson ($P$)
plus one vector meson ($V$), that is
\bea
x \delta\qbar^\sigma =\sum_{\lambda=0,\pm 1}
f^\lambda_{V B/N} \otimes \qbar^\sigma_{V_{\lambda}}
+\sum_{\lambda=0,\pm 1}
f^\lambda_{(V_1 V_2)B/N} \otimes \qbar^\sigma_{(V_1 V_2)_{\lambda}}
+
\sum_{\lambda=0}
f^0_{(P V)B/N} \otimes \qbar^\sigma_{(P V)_{0}}.
\label{xqbar3}
\eea
Observing that (see the discussion below)
\bea
\qbar^\ua_{(V_1 V_2)_{1}} = \qbar^\da_{(V_1 V_2)_{-1}}, &~~&
\qbar^\da_{(V_1 V_2)_{1}} = \qbar^\ua_{(V_1 V_2)_{-1}}, \nonumber\\
\qbar^\ua_{(V_1 V_2)_{0}} = \qbar^\da_{(V_1 V_2)_{0}}, &~~&
\qbar^\ua_{(P V)_{0}} \not= \qbar^\da_{(P V)_{0}},
\eea
and denoting
\bea
\Delta f_{(V_1 V_2)B/N} = f^1_{(V_1 V_2)B/N}-f^{-1}_{(V_1 V_2)B/N} \\
\Delta \qbar_{V_1 V_2} = \qbar^\ua_{(V_1 V_2)_{1}}-\qbar^\da_{(V_1 V_2)_{-1}}, ~~
\Delta \qbar_{(PV)_{0}} = \qbar^\ua_{(P V)_{0}}-\qbar^\da_{(P V)_{0}}
\eea
we have 
\bea 
x  (\Delta \qbar) &=& x\delta\qbar^\ua - x\delta\qbar^\da \nonumber \\
&=& \Delta f_{VB/N}\otimes\Delta\qbar_V
+\Delta f_{(V_1 V_2) B/N}\otimes \Delta \qbar_{V_1 V_2}
+\Delta f_{(P V) B/N}\otimes \Delta \qbar_{P V}.
\label{xqbar4}
\eea
The first term in Eq.~(\ref{xqbar4}) comes from the vector meson cloud,
which has been considered in \cite{RFriesS}.
The second and third terms, which are first included in this study,
result from the interference between terms with two different vector mesons 
($\rho,\, \omega$) and between terms with a vector meson ($\rho,\,\omega$) 
and a pseudoscalar meson ($\pi$) respectively.
Eq.~(\ref{xqbar4}) explicitly shows the existence of the interference contributions.
We would like to point out that the above interference terms do not contribute
to the unpolarized parton distributions due to the flavour-spin
structure of the $SU(6)$ wavefunction.
For example, the $\pi^+$-$\rho^+$ interference term contributes to $\dbarup$ and
$\dbardn$ with equal magnitude but opposite sign
(see the below expressions for the wave functions), so the result is zero
when the helicities are summed up.
The $SU(6)$ wavefunction also leads to a zero contribution to the polarised 
anti-quark distribution from $\pi$-$\eta$ interference terms.

The interference distributions ($\Delta\qbar_{\rho\omega}, \, \Delta\qbar_{\pi\rho}, \,
\Delta\qbar_{\pi,\omega}, \, q=u,d$)
do not have the same straightforward interpretation as the quark distributions.
However using the quark model with $SU(6)$ wavefunctions we can relate
these distributions to the polarised anti-quark distributions of the vector mesons.
In the quark model, the valence wavefunctions of the $\pi,\, \rho$ and $\omega$
mesons can be written as \cite{CloseBook},
\bea
|\pi^+\rangle &=& \frac{1}{\sqrt{2}}(\dbarup \udn-\dbardn \uup)
			\psi_\pi(x,k_\perp), \nonumber \\
|\pi^0\rangle &=& \frac{1}{2}(\ubarup \udn- \ubardn \uup
		-\dbarup \ddn+\dbardn \dup_)\psi_\pi(x,k_\perp), \nonumber \\
|\pi^-\rangle &=& \frac{1}{\sqrt{2}}(\ubarup \ddn-\ubardn \dup)
			\psi_\pi(x,k_\perp),\\
|\rho^+\rangle^{1,1}&=&\dbarup\uup \psi_\rho(x,k_\perp), \nonumber \\
|\rho^+\rangle^{1,0}&=&\frac{1}{\sqrt{2}}(\dbarup\udn+\dbardn\uup)
							\psi_\rho(x,k_\perp), \nonumber \\
|\rho^+\rangle^{1,-1}&=&\dbardn\udn \psi_\rho(x,k_\perp), \nonumber \\
|\rho^0\rangle^{1,1}&=&\frac{1}{\sqrt{2}}(\ubarup\uup-\dbarup\dup)
						\psi_\rho(x,k_\perp), \nonumber \\
|\rho^0\rangle^{1,0}&=&\frac{1}{2}(\ubarup\udn+\ubardn\uup
						-\dbarup\ddn-
\dbardn\dup)\psi_\rho(x,k_\perp), \nonumber \\
|\rho^0\rangle^{1,-1}&=&\frac{1}{\sqrt{2}}(\ubardn\udn-\dbardn\ddn)
						\psi_\rho(x,k_\perp), \nonumber \\
|\rho^-\rangle^{1,1}&=&\ubarup\dup \psi_\rho(x,k_\perp), \nonumber \\
|\rho^-\rangle^{1,0}&=&\frac{1}{\sqrt{2}}(\ubarup\ddn+\ubardn\dup)
							\psi_\rho(x,k_\perp), \nonumber \\
|\rho^-\rangle^{1,-1}&=&\ubardn\ddn \psi_\rho(x,k_\perp), \\
|\omega\rangle^{1,1}&=&\frac{1}{\sqrt{2}}(\ubarup\uup+\dbarup\dup)
						\psi_\omega(x,k_\perp), \nonumber \\
|\omega\rangle^{1,0}&=&\frac{1}{2}(\ubarup\udn+\ubardn\uup
			+\dbarup\ddn+\dbardn\dup)\psi_\omega(x,k_\perp), \nonumber \\
|\omega\rangle^{1,-1}&=&\frac{1}{\sqrt{2}}(\ubardn\udn+\dbardn\ddn)
						\psi_\omega(x,k_\perp),
\eea
where $\psi_{M}(x,k_\perp)$ is a two-body light-cone wave function.
The $\omega$ meson has been treated as an ideal mixture of an octet and a 
singlet.
Note that the distribution $\phi(x)=\int d^2 {\bf k}_\perp |\psi(x,k_\perp)|^2$
is not the ``true" parton distribution since only the lowest Fock state
is considered and the normalization condition is not satisfied ($\int_0^1\phi(x) < 1$).
Employing the above wave functions and assuming
$\psi_\pi(x,k_\perp)=\psi_\rho(x,k_\perp)=\psi_\omega(x,k_\perp)$,
we can obtain the following relations between the polarized anti-quark distributions
and the interference distributions,
\bea
\Delta\dbar_{\rho^+}&=&\Delta \ubar_{\rho^-}=
2\Delta\dbar_{\rho^0}=2\Delta \ubar_{\rho^0}=
2\Delta\dbar_{\omega}=2\Delta \ubar_{\omega}=\phi(x), \nonumber \\
\Delta\dbar_{\rho^0\omega}&=&-\Delta\ubar_{\rho^0\omega}=
-\frac{1}{2}\phi(x),\nonumber \\
\Delta\dbar_{(\pi^+\rho^+)_{0}}&=&\Delta\ubar_{(\pi^-\rho^-)_{0}}
=2\Delta\dbar_{(\pi^0\rho^0)_{0}}=2\Delta\ubar_{(\pi^0\rho^0)_{0}}=\phi(x), 
\nonumber \\
\Delta\dbar_{(\pi^0\omega)_{0}}&=&-\Delta\ubar_{(\pi^0\omega)_{0}}=
-\frac{1}{2}\phi(x).
\label{pdrelations}
\eea
Although the above relations are derived from the quark model
and by considering only the lowest Fock states,
we will assume they hold for the full wavefunction.
Thus the distribution $\phi(x)$ can be replaced with the polarized
parton distribution
$\Delta v_\rho=\Delta \dbar_{\rho^+}=\Delta \ubar_{\rho^-}$ which,
in principle, can be measured experimentally. We adopt two prescriptions 
to obtain the $\Delta v_\rho$  distribution: (i) employing the MIT bag model 
and (ii) adopting the ansatz used in \cite{RFriesS}, {\it i.e.} relating 
it to the valence quark distribution of the $\pi$ meson inspired by the
lattice calculation of the first moments of the polarized and unpolarized
parton distributions of the $\rho$ meson.

We will consider the fluctuations $p \ra N \pi, N \rho, N \omega$ and 
$p \ra \Delta \pi, \Delta \rho$.
We neglect the fluctuation $p \ra \Delta \omega$ as this fluctuation is 
forbidden by isospin.
The following relations exist for the fluctuation functions 
\cite{WMelnitchoukST},
\bea
\Delta f_{\rho^+n/p}=2\Delta f_{\rho^0 p/p}=\frac{2}{3}\Delta f_{\rho N/N},
\nonumber \\
\Delta f_{\rho^-\Delta^{++}/p}=\frac{3}{2}\Delta f_{\rho^0\Delta^+/p }
=3\Delta f_{\rho^+\Delta^0/p}=\frac{1}{2}\Delta f_{\rho\Delta/N},
\nonumber \\
f_{(\pi^+\rho^+)n/p}=2 f_{(\pi^0\rho^0)p/p}=f_{(\pi \rho)N/N},
\nonumber \\
f_{(\pi^-\rho^-)\Delta^{++}/p}=\frac{3}{2}f_{(\pi^0\rho^0)\Delta^+/p}
=3 f_{(\pi^+\rho^+)\Delta^0/p}=\frac{1}{2}f_{(\pi \rho)\Delta/N}.
\label{ffrelations}
\eea
Using Eqs.~(\ref{pdrelations}) and (\ref{ffrelations}) we can obtain from
Eq.~(\ref{xqbar4}),
\bea
x(\Delta\dbar -\Delta \ubar)&=&~~[\frac{2}{3} \Delta f_{\rho N/ N}
-\frac{1}{3} \Delta f_{\rho \Delta/ N }] \otimes\Delta v_\rho \nonumber\\
&~~&+[-\Delta f_{(\rho^0\omega)p/p}+\frac{2}{3} f^0_{(\pi\rho)N/N}
-\frac{1}{3} f^0_{(\pi\rho)\Delta/N} - f^0_{(\pi^0\omega)p/p}]
\otimes\Delta v_\rho \nonumber \\
&=&\Delta f_\rho\otimes\Delta v_\rho +
	\Delta f_{int}\otimes\Delta v_\rho.
\label{xDeltadu}
\eea
The first term is the same as the result given in \cite{RFriesS}. We note that there 
are no contributions directly from the $\omega$ meson due to its charge structure.
The second term is the interference contribution.

Now we turn to the calculation of the fluctuation functions.
The fluctuation $N \ra MB$ is described by the effective interaction Lagrangians 
\cite{HHoltmannSS},
\bea
{\cal L}_{NN\pi}&=&i g_{NN\pi} {\bar N}\gamma_5\pi N \nonumber \\
{\cal L}_{N\Delta \pi}&=&f_{N\Delta \pi} {\bar N} \partial_\mu \pi\Delta^\mu + 
\mbox{h.c.},
\nonumber \\
{\cal L}_{NNV}&=&g_{NNV} {\bar N} \gamma_\mu\theta^\mu N
			     +f_{NNV} {\bar N} \sigma_{\mu\nu}
		N (\partial^\mu\theta^\nu-\partial^\nu\theta^\mu)
\nonumber \\
{\cal L}_{N\Delta\rho}&=&f_{N\Delta \rho} i {\bar N} \gamma_5 \gamma_\mu \Delta_\nu	
	(\partial^\mu\theta^\nu-\partial^\nu\theta^\mu) + \mbox{h.c.},
\label{langragians}
\eea
where $N$ is a spin-1/2 field, $\Delta$ a spin-3/2 field of Rarita-Schwinger 
form,
$\pi$ a pseudoscalar field, and $\theta$ a vector field.
The coupling constants are taken to be \cite{HHoltmannSS,PReport},
\bea
g^2_{NN\pi}/4\pi=13.6, & \nonumber \\
g^2_{NN\rho}/4\pi=0.84, & f_{NN\rho}/g_{NN\rho}=6.1/4m_N, \nonumber \\
g^2_{NN\omega}/4\pi=8.1, & f_{NN\omega}/g_{NN\omega}=0, \nonumber \\
f^2_{N\Delta\pi}/4\pi=12.3~{\rm GeV}^{-2}, & ~~~
f^2_{N\Delta\rho}/4\pi=34.5~{\rm GeV}^{-2}.
\label{couplingconstant}
\eea
The amplitudes $\phi^{\lambda \lambda^\prime}_{MB}(y,k_\perp^2)$ which essentially
determine the fluctuation function (see Eq.~(\ref{ff})) are calculated by using 
time-ordered perturbation theory in the infinite momentum frame,
\bea
\phi^{\lambda \lambda^\prime}_{MB}(y,k_\perp^2)
=\frac{1}{2\pi\sqrt{y(1-y)}}
  \frac{\sqrt{m_N m_B}
       V_{IMF}^{\lambda\lambda^\prime}(y,k_\perp^2) \, G_{MB}(y,k_\perp^2)}
       {m_N^2-m^2_{MB}(y,k_\perp^2)},
\eea
where $m_{MB}^2$ is the invariant mass squared of the $MB$ Fock state,
\bea
m_{MB}^2(y,k_\perp^2)=\frac{m_M^2+k_\perp^2}{y}
+\frac{m_B^2+k_\perp^2}{1-y}.
\eea
As usual a phenomenological vertex form factor, $G_{MB}(y,k_\perp^2)$
is introduced to describe
the unknown dynamics of the fluctuation $N\ra MB$. Here we adopt 
the exponential form,
\bea
G_{MB}(y,k_\perp^2)=exp\left[\frac{m_N^2-m_{MB}(y,k_\perp^2)}
{2\Lambda^2}\right],
\eea
where $\Lambda$ is a cut-off parameter.
We adopt $\Lambda_{oct}=1.08$ GeV and and $\Lambda_{dec}=0.98$ GeV
for the fluctuations involving the octet and decuplet baryons respectively 
\cite{HHoltmannSS}.
This form factor satisfies the relation
$G_{MB}(y,k_\perp^2)=G_{BM}(1-y,k_\perp^2)$.

We note that there are two prescriptions for calculating the vertex functions 
$V_{IMF}$, depending on the manner in which the meson energy is treated. 
In this work we follow the prescription used in reference \cite{HHoltmannSS}
{\it i.e.} the meson energy in the vertex is $E_N-E_B$,
which is called method (B) in reference \cite{RFriesS}.
The expressions for the various fluctuation functions in Eq.~(\ref{xDeltadu}) are
given in the Appendix. These expressions agree with the vertex functions given
in \cite{HHoltmannSS}, but differ from those given in the appendix of 
\cite{RFriesS} in the following ways:
1) The terms proportional to $f_{NN\rho} \, g_{NN\rho}$ in $f_{\rho N/N}^\lambda(y)$
have the opposite sign;
2) Our results for $f_{\rho \Delta/N}^\lambda (y)$
agree with the results of method (A) in \cite{RFriesS}
(Eqs.~(24) and (26) in the appendix of \cite{RFriesS}), rather than
the results of method (B).
Finally we note that adopting the 
alternative method (method (A) of \cite{RFriesS}) leads to somewhat 
smaller values of $(\Delta\dbar -\Delta \ubar)$, but does not change our 
conclusions significantly.

There is little experimental information on the parton distributions of the 
vector meson. Although it is common practice to set the unpolarized parton 
distribution of the $\rho$ meson the same as that of the pion,
the study of the polarized parton distribution
of the $\rho$ meson is lacking both in experiment and theory.
The lattice calculation \cite{Lattice} finds that the polarization of the $\rho$ meson 
is  about $60\%$. So the ansatz $\Delta v_\rho(x)=0.6  v_\pi(x)$ was used
in \cite{RFriesS}.
We note that the lattice prediction of $60\%$ polarization is for the ratio of the first
moments of the polarized and unpolarized parton distributions,
{\it i.e.} $\int_0^1 \Delta v_\rho(x)=0.6 \int_0^1 v_\rho(x)$ and it is quite possible that 
the $x$-dependence of the polarized parton distribution may be different
from that of the unpolarized one.

As an alternative hypothesis for the $x$-dependence of the polarized parton 
distribution, we employ a non-perturbative model of hadrons -- 
the MIT bag model \cite{MIT}.
The bag model has been shown to be a useful tool in the study of the 
non-perturbative structure of hadrons (e.g., mass spectrum, parton distribution).
The theoretical calculations \cite{ASignalT,ASchreiberTL,ASchreiberST} of the 
parton distributions of the nucleon, including meson cloud contributions,
can give results consistent with the experimental data.
An interesting aspect of the bag model calculation is that it can be generalised to 
provide useful information on the parton distributions of the other hadrons. 
The parton distributions for both polarized and unpolarized octet and decuplet
baryons have been calculated in the bag model \cite{CBorosT}.
However most present bag model calculations for
the parton distributions are for the baryons.
There has been no attempt in the bag model to calculate
the parton distributions of the mesons. 
This is due, at least in part, to the lack of
experimental data on the parton distributions of
the mesons\footnote{At present only the parameterization for
the parton distributions of the pion has been extracted experimentally 
\cite{SMRSPion}.}.
While the bag model is probably not very applicable to the pion, it does 
describe the rest of the pseudoscalar nonet and the vector octet reasonably well. 
So adapting the methods used to calculate baryon parton distributions to the meson 
sector should give a useful approximation to the parton distributions of 
the mesons, in particular the $\rho$ meson.

Adapting the argument of reference \cite{ASchreiberST} we obtain the expression 
for the quark distribution function in a $\rho$ meson, where we include only 
one-quark intermediate states
\bea
q_{\rho,f}^{\ua \da}(x) = 
\frac{M_{\rho}}{(2\pi)^{3}} \sum_{m} \langle \mu | P_{f,m}| \mu \rangle 
\int d{\bf p}_{n} \frac{|\phi_{1}({\bf p}_{n})|^{2}}{|\phi_{2}({\bf 0})|^{2}} 
\delta(M_{\rho}(1-x) - p_{n}^{+}) |\tilde{\Psi}^{\ua \da}_{+,f}({\bf p}_{n})|^{2}.
\label{mit2q}
\eea
Here + components of momenta are defined by $p^+ = p^0 + p^3$, 
${\bf p}_{n}$ is the 3-momentum of the intermediate state, 
$\tilde{\Psi}$ is the Fourier transform of the MIT bag ground state wavefunction
$\Psi({\bf r})$, and $\phi_{m}({\bf p})$ is the Fourier transform of the Hill-Wheeler 
overlap function between $m$-quark bag states:
\bea
|\phi_{m}({\bf p})|^{2} = \int d{\bf R} e^{-i{\bf p \cdot R}}
\left[ \int d{\bf r} \Psi^{\dagger}({\bf r-R}) \Psi({\bf r}) \right]^{m}.
\eea
The $\phi$ functions arise through the use of the Peierls-Yoccoz projection to 
form momentum eigenstates from the initial and intermediate bag states. 
The matrix element $\langle \mu | P_{f,m}| \mu \rangle$ appearing in  
eqn.~(\ref{mit2q}) is the matrix element of the projection operator 
$P_{f,m}$ onto the required flavour $f$ and helicity $m$ for the $SU(6)$ 
spin-flavour wavefunction $| \mu \rangle$ of the $\rho$ meson.

\section{Results and Discussion}

We first fix the parameters of the MIT bag model calculation by fitting the calculated
unpolarized parton distribution of the $\rho$ meson to the Gluck-Reya-Schienbein
parameterization (GRS99) \cite{GRS99} for the valence parton distribution
of the pion\footnote{As usual we take the unpolarized pion and $\rho$ valence 
distributions to be the same.}, 
which is essentially fixed by the $\pi$-$N$ Drell-Yan data.
For the parton distributions of the nucleon, the bag model calculations
with only two-quark intermediate states are usually smaller than the data
in the small-$x$ region and do not satisfy the normalization condition of the
probability, {\it i.e.} $P_2 < 1$ rather than $P_2=1$
\cite{ASignalT,ASchreiberTL,ASchreiberST}.
It is necessary to include intermediate states with three quarks and one 
anti-quark, arising from the action of the field operator ($\psi = b + d^{\dagger}$)
on the three-quark bag state. This allows the normalization condition to be 
satisfied.
Such contributions from multi-quark intermediate states can be well parametrised 
by the form 
$f_4(x)=N_4(1-x)^7$
consistent with the Drell-Yan-West relation \cite{ASchreiberST}.
For simplicity we will employ a similar function form 
$f_3(x) = N_3 (1-x)^5$ 
to parametrise the three-quark  (and two-quark plus one-antiquark) intermediate 
state contributions to the unpolarized parton distributions of the $\rho$ meson.
The value of $N_3$ is determined from the normalization condition
$P_1+P_3=1$,
where $P_1$ and $P_3$ are the probabilities of the one-quark and three-quark
intermediate states respectively.
The parameters needing to be fixed are the radius of the bag $R$, the mass
of the one-quark intermediate state $m_1$, and the low momentum scale $\mu^2$, 
at which the model is supposed to be valid.
The next-to-leading-order GRS99 parameterization is given at a scale,
$\mu_{NLO}^2=0.40$~GeV$^2$,
\bea
v_{NLO}^\pi(x,\mu^2_{NLO})=0.696x^{-0.447}(1-x)^{0.426},
\eea
where $v^\pi=u_v^{\pi^+}=\dbar_v^{\pi^+}$.
Both the GRS99 parton distributions and our calculations are evolved to
the scale $Q^2=4$~GeV$^2$ and the results are shown in Fig.~2.
The thin dashed curve is the GRS99
parameterization, and the thick dashed curve is the bag model calculation.
A good agreement in the small and  intermediate
$x$ region is found for $R=0.7$ fm, $m_1=0.55m_\rho$, $N_3=1.68$
and $\mu^2=0.23$~GeV$^2$.

Having fixed the parameters we calculate the polarized parton distribution of
the $\rho$, $x \Delta v_{\rho}^{MIT}(x)$. The result is presented in Fig.~2 as the 
thick solid curve.
The first moment of $\Delta v_{\rho}^{MIT}(x)$ is found to be about $0.60$ at 
$Q^2=4$~GeV$^2$, which is in agreement with the lattice value of $0.60$.
For comparision, the distribution $ 0.6\,xv_\pi(x)$, which could be set as the 
polarized parton distribution according to the ansatz used in \cite{RFriesS}, is 
also shown in Fig.~2 as the thin solid curve.
It can be seen that the distribution $ 0.6\,xv_\pi(x)$ is smaller than the bag model 
calculation $x \Delta v_{\rho}^{MIT}(x)$ in the intermediate $x$ region,
although both distributions satisfy the same normalization condition.
Also the bag model polarized parton distribution has a different
$x$-dependence from the unpolarized distribution.

We show the polarized fluctuation functions of the $\rho$ meson
($\Delta f_\rho$ in Eq.(\ref{xDeltadu}),  the solid curve) and
the interference terms ($\Delta f_{int}$ in Eq.(\ref{xDeltadu}), the dashed curve)
in Fig.~3.
It can be seen that the maximum of $\Delta f_{int}$ is about $40\%$ that of
$\Delta f_\rho$ and that $\Delta f_{int}$ changes sign from positive to negative
at about $y=0.6$.
So the contribution to $\Delta \dbar -\Delta \ubar$ from the interference terms
is not negligible.

As we have discussed in the last section, our expressions for $f^\lambda_{\rho N/N}$
and $f^\lambda_{\rho \Delta/N}$ are different from those given in \cite{RFriesS}.
We show the numerical difference in Fig.~4 where
the fluctuation functions $\Delta f_{\rho NN}$ 
(dashed curves), $\Delta f_{\rho \Delta N}$ (dotted curves), and
$\Delta f_\rho=\frac{2}{3}\Delta f_{\rho NN}- \frac{1}{3} \Delta f_{\rho \Delta N}$
(solid curves) which enter directly in the calculation of
$\Delta \dbar - \Delta \ubar$ (see Eq.~(\ref{xDeltadu})) are plotted.
The thick curves are our results while the thin curves are from \cite{RFriesS}. 
In each case the cut-off parameter in the form factor has been set to the same 
value $\Lambda_{oct}=\Lambda_{dec}=0.85$~GeV.
The difference is  about 50\%.

We calculate the flavour asymmetry of the polarized anti-quark distributions
employing both the bag model distribution $x \Delta v_{\rho}^{MIT}(x)$  
and $ 0.6\, xv_\pi(x)$ for the polarized valence parton distribution of the 
$\rho$ meson.
The results are shown in Fig.~5.
The solid curves are the predictions using $x \Delta v_{\rho}^{MIT}(x)$, while
the dashed curves are obtained by using $ 0.6\, xv_\pi(x)$.
The thin curves are the results without interference terms while the thick
curves are the results with interference terms.
We see that the interference effect mildly increases the predictions for the 
flavour asymmetry, and pushes the curves towards the small $x$ region 
due to  $\Delta f_{inter}$ being peaked at smaller $y$ ($y_{max}\simeq 0.3$) 
than the $\Delta f_\rho$ ($y_{max}\simeq0.60$). 
For the calculations with $x \Delta v_{\rho}^{MIT}(x)$, the asymmetry with 
interference terms included exhibits a maximum at $x\simeq 0.12$ while the 
asymmetry without interference terms exhibits a maximum at $x\simeq 0.18$.
Also it can be seen that the calculations with $x \Delta v_{\rho}^{MIT}(x)$ 
(the solid curves) are larger than that with $ 0.6\, xv_\pi(x)$ 
(the dashed curves) in the intermediate $x$ region, and have their maxima 
at larger $x$.

The integral 
\bea
I_{\Delta} & = & \int_{0}^{1} dx [\Delta \dbar(x) -\Delta \ubar(x)] \nonumber \\
& = & \int_{0}^{1} dx \Delta v_{\rho}(x) 
\int_{0}^{1} dy [\Delta f_{\rho}(y) + \Delta f_{int}(y)]
\eea 
will be the same for both models for the polarized parton distribution of the 
$\rho$ as they have the same first moment for the polarized distribution.
We find the integral to be $0.023$ ($0.031$) without (with) the interference 
terms.
The interference effect increases the integral by about $30\%$.

Using a softer form factor for the octet contributions, as suggested by the fit of
$\dbar(x) - \ubar(x)$ at large $x$ \cite{WMelnitchoukST},
will lower the value of the integral $I_{\Delta}$.
For example, taking $\Lambda_{oct} = 0.80$ GeV and $\Lambda_{dec} = 1.0$~GeV  
consistent with the analysis of \cite{WMelnitchoukST},
the integral decreases to $-1.1\times 10^{-4}$ ($+5.7\times 10^{-3}$)
without (with) the interference terms.
In this case the flavour asymmetry between $\dbar$ and $\ubar$
is very small.
The reason is that the fluctuation $p \ra \Delta \rho$ gives a negative contribution
to the integral $I_{\Delta}$ and this fluctuation is greatly emphasized
for the above cut-off parameters\footnote{
The probabilities for $p \ra N \rho$ and $p \ra \Delta \rho$ are $0.012$
and $0.042$ respectively, while the probabilities are $0.189$ and $0.034$
for the parameter values $\Lambda_{oct} = 1.08$ GeV and $\Lambda_{dec} = 0.98$~GeV.}.
The prediction for the integral $I_{\Delta}$ has a strong dependence on the cut-off
parameters $\Lambda_{oct}$ and $\Lambda_{dec}$. For example,
the results calculated including interference terms vary from $0.0043$ to $0.033$ 
for the cut-off parameters changing from $\Lambda_{oct}=\Lambda_{dec}=0.8$~GeV 
to $1.10$~GeV.
Clearly these values obtained using the meson cloud model are very different 
from those obtained using the chiral quark-soliton model \cite{LargeNC2}
which have a magnitude of around $0.3$.
It is interesting that both models agree well with the experimental data 
for the unpolarized asymmetry, yet predict very different results for the 
polarized asymmetry. As the magnitude of the predicted polarized asymmetry 
appears to be fairly natural in each of these models, experimental data will 
provide a valuable test of these models, and give insight into the relation 
between helicity dependent and helicity independent observables in 
quark models.

We do not find that the $\pi - \rho$ interference terms can be sizeable, which 
appears to be in disagreement with the conclusions of Boreskov and Kaidalov
\cite{Regge}. 
The main reason for this disagreement is that we do not here 
consider interference terms where the $\rho$ meson has non-zero helicity. 
This is because any such terms only contribute to amplitudes in the virtual 
Compton scattering which have a spin-flip between the incoming and outgoing 
proton states. 
These spin-flip amplitudes in turn only contribute to the 
cross-section $\sigma_{I}$, which is the interference between transverse and 
longitudinal polarisations of the virtual photon \cite{Robertsbook}
\bea
\sigma_{I} &\propto & \frac{\sqrt{Q^2}}{M^2} \left[ G_{1} + \frac{\nu}{M}G_{2}\right] 
\nonumber \\
&\rightarrow & \frac{\sqrt{Q^2}}{M \nu} \left[ g_{1}(x) + g_{2}(x) \right] \,
\mbox{in the Bjorken limit.}
\eea
So any interference terms involving non-zero helicity $\rho$ mesons can be 
expected to decrease as $1/\sqrt{Q^2}$ as $Q^2$ gets large. 
Using the operator product expansion shows that the relevant operators 
are all twist 3 or higher. 
As the experimental data for $g_1(x)$ for both the proton and the neutron 
show no marked $Q^2$ dependence, we conclude that these higher twist 
contributions are not relevant at the experimental scales.

\section{Pauli blocking contributions to the flavour asymmetry}
We have not so far considered any contribution to the asymmetry arising 
from `Pauli blocking' effects \cite{WMelnitchoukST,ASignalT,FieldF}. 
In a model such as the bag model, where the valence quarks are confined 
by a scalar field, the vacuum inside a hadron is different from the vacuum
outside. 
This manifests itself as a distortion in the negative energy Dirac sea, which
is full for the outside (or free) vacuum, whereas there will be empty states in 
the Dirac sea of the bag.
To an external probe this change in vacuum structure appears as an 
intrinsic, non-perturbative sea of $q\bar{q}$ pairs \cite{DunneTT}. 
This change in the vacuum is similar to the change in the Fermi-Dirac distribution 
when the temperature is raised above absolute zero. 
Now when a quark is put into the ground state of the bag it wil have the effect of 
filling some of the empty negative energy states in the sea of the bag vacuum. 
The reason for this is that the ground state wavefunction can be written as a 
wavepacket in terms of plane wave states of positive and negative energy, with the 
energy distribution of the wavepacket centred at the ground state energy eigenvalue, 
but with non-zero contributions from negative energy plane waves. 
Hence the presence of a quark in the bag ground state lowers the probability of 
a negative energy state being empty, which is the same as lowering the probability 
of finding a positive energy antiquark. 
As the proton consists of two up quarks and one down quark, the probability of 
finding a $\bar{u}$ antiquark is reduced more than the probability of finding a 
$\bar{d}$ antiquark {\it i.e.} $\bar{d} > \bar{u}$. 
The analysis of ref \cite{WMelnitchoukST} showed that, in the context of the
meson cloud model, about 50\% of the observed $\bar{d} - \bar{u}$ asymmetry 
may be due to Pauli blocking. 

When we include spin in the analysis of Pauli blocking, we find that putting a 
spin up quark into the bag ground state has the effect of filling some of the 
negative energy spin up quark states in the bag vacuum, which is equivalent to 
lowering the probability of finding a positive energy spin down antiquark. 
As the $SU(6)$ wavefunction of the spin up proton is dominated by terms with the 
two up quarks having spin parallel to the proton spin and the down quark having 
spin anti-parallel, Pauli blocking predicts that the probabilities of finding spin 
down $\bar{u}$ antiquarks and spin up $\bar{d}$ antiquarks are reduced {\it i.e.} 
$\bar{u}^{\ua} > \bar{u}^{\da}, \; \bar{d}^{\da} > \bar{d}^{\ua}$ or 
$\Delta \bar{u}(x) \geq 0, \; \Delta \bar{d}(x) \leq 0.$

We can also estimate the contribution of the Pauli blocking effect to the 
polarized asymmetry, again using the Adelaide group's argument for calculating 
parton distributions in the bag model. 
In the parton model antiquark distribution functions are given by
\bea
\bar{q}^{\ua \da}(x) = p^{+} \sum_{n} \delta (p^{+}(1-x)-p_{n}^{+}) \left| \langle n | 
\frac{1}{2}(1 \mp \gamma^{5})\Psi^{\dagger}_{+}(0) | p, s \rangle \right|^{2}.
\eea
The appropriate intermediate state $|n \rangle$ consists of four quarks.
If we assume the $SU(6)$ wavefunction for the proton with spin $+1/2$, and 
insert an additional quark only into the radial ground state, then we have 
the following matrix elements for the projection operators onto spin and 
flavour \cite{ASchreiberST} 
\bea
\langle \mu |P^{(q)}_{u, +1/2}| \mu \rangle = \frac{4}{3}, \; & &
\langle \mu |P^{(q)}_{u, -1/2}| \mu \rangle = \frac{8}{3},  \nonumber \\
\langle \mu |P^{(q)}_{d, +1/2}| \mu \rangle = \frac{8}{3}, \; & &
\langle \mu |P^{(q)}_{d, -1/2}| \mu \rangle = \frac{7}{3}.
\eea
We have ignored any effects of spin-spin coupling in the intermediate state.
Following the argument of the Adelaide group for calculating the parton distributions
we can then write the antiquark distributions as 
\bea
\bar{u}^{\ua \da} & = & 2 F_{(4)}(x) \pm \frac{2}{3} G_{(4)}(x), \nonumber \\
\bar{d}^{\ua \da} & = & \frac{5}{2} F_{(4)}(x) \mp \frac{1}{6} G_{(4)}(x), \label{eq:qbardist}
\eea
where $F_{(4)}(x)$ and $G_{(4)}(x)$ are the spin independent and spin dependent 
kinematic integrals over the momentum of the intermediate four quark state.
The sea asymmetries can then be expressed as 
\bea
\bar{d}(x) - \bar{u}(x) &=&  F_{(4)}(x), \nonumber \\
\Delta \bar{d}(x) - \Delta \bar{u}(x) &=& -\frac{5}{3} G_{(4)}(x).
\eea
As $G_{(4)}(x)$ is positive at all $x$, Pauli blocking gives a negative 
contribution to the spin dependent flavour asymmetry in the sea, whereas the 
meson cloud contribution tended to be positive.
Also noting that as $F_{(4)}(x) \geq G_{(4)}(x)$, we can integrate over all $x$ 
and then obtain an upper limit for the size of the Pauli blocking contribution to the spin 
dependent asymmetry in terms of the contribution to the spin independent asymmetry:
\bea
-\int_{0}^{1} dx [ \Delta \bar{d}(x) - \Delta \bar{u}(x) ] \leq 
\frac{5}{3} \int_{0}^{1} dx [ \bar{d}(x) - \bar{u}(x) ]. \label{pbsums}
\eea
As an estimate for the integral on the rhs of eqn. (\ref{pbsums}) we may 
use the value of $0.07$ given by the analsis of reference \cite{WMelnitchoukST}.
This then gives an upper limit of about $0.12$ for the magnitude of the integral 
over the polarized asymmetry. 
In the bag model, the ratio $G_{(4)}(x) / F_{(4)}(x)$ varies from about $0.8$ at 
low $x$ to unity at large $x$, which gives us a value of about $-0.09$ for the 
integrated polarized asymmetry.
While these values are calculated at some scale appropriate to the bag model, 
the values of the integrals are not much affected by evolution up to experimental 
scales, so we expect the relation between polarized and unpolarized sea 
asymmetries to be approximately correct at all scales.
The value of the Pauli blocking contribution to the integrated polarized asymmetry 
is much larger than that we have calculated in the meson cloud model, 
in contrast to approximate equality in the unpolarized case. 
Thus the experimental observation of any asymmetry in the polarized sea 
distributions is much more a test of the Pauli blocking hypothesis than of the 
meson cloud model.

The Bjorken sum rule may be written
\bea
\int_{0}^{1} dx [g_{1}^{p}(x) - g_{1}^{n}(x)] & = & 
\frac{1}{6}\int_{0}^{1} dx [(\Delta u(x) - \Delta d(x)) + 
(\Delta \bar{u}(x) - \Delta \bar{d}(x))] \nonumber \\
& = & \frac{1}{6} \left| \frac{g_{A}}{g_{V}} \right| 
\left(1 - \frac{\alpha_{s}}{\pi} \right).
\eea
We estimate that the contribution to the sum rule from Pauli blocking plus meson 
cloud effects is about 5-10\% of the value of the sum rule.

We note that the Dortmund group have recently \cite{Dort} analyzed the polarized 
sea asymmetry also using a Pauli blocking type ansatz, and found a value around 
$-0.3$ for the integrated asymmetry. This would correspond to a contribution of 
around 20\% to the Bjorken sum rule. 
The Dortmund analysis was based on the proposed relation between polarized 
distributions:
\bea
\frac{\Delta \bar{d}(x)}{\Delta \bar{u}(x)} = \frac{\Delta u(x)}{\Delta d(x)}.
\label{dortpol}
\eea 
This relation is not obeyed by the distributions in our analysis. 
The reason for this is that Pauli blocking should most affect the $\Delta \bar{u}$ 
distribution rather than the $\Delta \bar{d}$ distribution (starting from an 
assumed $SU(6)$ value of 0), and hence the lhs of the relation (\ref{dortpol}) has 
magnitude less than one, while the magnitude of the rhs is greater than one. 

\section{Summary}
The meson cloud model is very successful in explaining  the
flavour asymmetry of the unpolarized parton distributions in the nucleon sea.
In this paper, we have calculated the flavour asymmetry for the polarized 
anti-quark distributions of the nucleon.
We have included the contributions from both the vector meson cloud and
the interference terms between pseudoscalar and vector mesons.
We have used two prescriptions to describe the polarized valance quark distribution
of the $\rho$ meson -- (i) calculating it in bag model and
(ii) employing the ansatz given in \cite{RFriesS} to relate it to the unpolarized
quark distribution of the $\pi$ meson.
Our calculations show that the interference effect mildly increases the prediction
for $\Delta \dbar(x)-\Delta \dbar(x)$ in the intermediate $x$ region.
We have also discussed the effect of `Pauli blocking' on the asymmetry, and have 
seen that this effect gives a larger contribution to the asymmetry than meson 
cloud effects, in contrast to the unpolarized case.

\section*{Acknowledgments}
We would like to thank Tony Thomas, Wally Melnitchouk, Andreas Schreiber and Kazuo 
Tsushima for a number of useful discussions, especially in regard to the sign of 
Pauli blocking contributions to the polarized sea asymmetry.
Part of this work was carried out while the authors were guests at the 
Centre for the Subatomic Structure of Matter at the University of Adelaide. 
This work was partially supported by the Science and Technology Postdoctoral
Fellowship of the Foundation for Research Science and Technology, and the 
Marsden Fund of the Royal Society of New Zealand.

\section*{Appendix}
We give here expressions for the helicity dependent fluctuation functions appear
in Eq.~(\ref{xDeltadu}), where the superscript $1$ ($-1$) is the vector meson helicity.
\bea
\Delta f_{\rho N/N}&=&f^1_{\rho N/N}-f^{-1}_{\rho N/N}
\eea
\bea
f^1_{\rho N/N}(y)&=&
\frac{3}{8\,{\pi }^2\,{\left( -1 + y \right) }^2\,y^3}
\int_0^\infty 
\frac{d k_\perp^2 G_{\rho N}^2(y,k_\perp^2)}
     {[m_N^2-m_{\rho N}^2(y,k_\perp^2)]^2 } \nonumber\\
& &\left\{
	g_{NN\rho}^2\,
       \left( k^2 + m_N^2\,y^4 \right) 
       \right. \nonumber \\
& & + 4\,f_{NN\rho}^2\,
       \left( k^4 + 5\,k^2\,m_N^2\,y^2 + 
         4\,m_N^4\,y^4 \right) \nonumber \\
& & \left.
	- 4\,f_{NN\rho}\,g_{NN\rho}\,m_N\,y\,
       \left[ 2\,m_N^2\,y^3 + 
         k^2\,\left( 1 + y \right)  \right] \right\} \\ \nonumber \\
f^{-1}_{\rho N/N}(y)&=&
\frac{3}{8\,{\pi }^2\,{\left( -1 + y \right) }^2\,y^3}
\int_0^\infty 
\frac{d k_\perp^2 G_{\rho N}^2(y,k_\perp^2)}
     {[m_N^2-m_{\rho N}^2(y,k_\perp^2)]^2} \nonumber\\
& &\left\{
	k^2\,\left[ g_{NN\rho}^2\,
       {\left( -1 + y \right) }^2 - 
      4\,f_{NN\rho}\,g_{NN\rho}\,m_N\,
       \left( -1 + y \right) \,y \right. \right. \nonumber \\
& & \left. \left. 
     + 4\,f_{NN\rho}^2\,
       \left( k^2 + m_N^2\,y^2 \right)  \right] \right\}
\eea

\bea
\Delta f_{\rho \Delta/ N}(y)&=&f^1_{\rho \Delta/ N}(y)-
								f^{-1}_{\rho \Delta/ N}(y)
\eea
\bea
f^1_{\rho \Delta/ N}(y)&=&
\frac{f_{N\Delta \rho}^2}
{24\, m_\Delta^2 {\pi }^2\,{\left( -1 + y \right) }^4\,y^3}
\int_0^\infty 
\frac{d k_\perp^2 G_{\rho \Delta}^2(y,k_\perp^2)}
     {[m_N^2-m_{\rho\Delta}^2(y,k_\perp^2)]^2} \nonumber\\
& &\left\{
     k^6 + k^4\,m_\Delta^2\,
       \left( 3 - 4\,y + 4\,y^2 \right)  \right. \nonumber \\
 & &    +  k^2\,m_\Delta\,y\,
       \left[ -4\,m_N\,m_\rho^2\,
          \left( -1 + y \right) ^3 + 
         m_\Delta^3\,y\,
          \left( 4 - 4\,y + 3\,y^2 \right) \right] \nonumber \\
& &      + \left. \left[ m_N\,m_\rho^2\,
           \left( -1 + y \right)^3 + 
           m_\Delta^3\,y^2 \right]^2 \right\} \\ \nonumber \\
f^{-1}_{\rho \Delta/ N}(y)&=&
\frac{f_{N\Delta \rho}^2}
{24\, m_\Delta^2 {\pi }^2\,{\left( -1 + y \right) }^2\,y^3}
\int_0^\infty 
\frac{d k_\perp^2 G_{\rho \Delta}^2(y,k_\perp^2)}
     {[m_N^2-m_{\rho\Delta}^2(y,k_\perp^2)]^2} \nonumber\\
& &\left\{
	  k^4\,m_N^2 + 
      k^2\,\left[ m_\rho^2\,
            \left( -1 + y \right)  + 
           2\,m_\Delta \,m_N\,y \right]^2 \right. \nonumber \\
& &   + \left.
       3\,m_\Delta^2\,
       \left[ m_\rho^2\,\left( -1 + y \right)  + 
           m_\Delta \,m_N\,y^2 \right]^2 \right\}
\eea
\bea
\Delta f_{(\rho^0 \omega) p/ p}=f^1_{(\rho^0 \omega) p/ p}
	-f^{-1}_{(\rho^0 \omega) p/ p}
\eea
\bea
f^1_{(\rho^0 \omega) p/p}(y)&=&
\frac{3 \, g_{NN \omega}}{8\,{\pi }^2\,{\left( -1 + y \right) }^2\,y^3}
\int_0^\infty 
\frac{d k_\perp^2 G_{\rho N}(y,k_\perp^2) G_{\omega N}(y,k_\perp^2)}
     {[m_N^2-m_{\rho N}^2(y,k_\perp^2)]
      [m_N^2-m_{\omega N}^2(y,k_\perp^2)]}
      \nonumber\\
& & \left\{
  g_{NN\rho}\,
       \left( k^2 + m_N^2\,y^4 \right)
      -2\, f_{NN \rho}\,m_N\,y\,
       \left[ 2\,m_N^2\,y^3 + 
         k^2\,\left( 1 + y \right)  \right]
         \right\}
         \\ \nonumber \\
f^{-1}_{(\rho^0 \omega) p/p}(y)&=&
\frac{3	\, g_{NN \omega}}{8\,{\pi }^2\,{\left( -1 + y \right) }\,y^3}
\int_0^\infty 
\frac{d k_\perp^2 G_{\rho N}(y,k_\perp^2) G_{\omega N}(y,k_\perp^2)}
     {[m_N^2-m_{\rho N}^2(y,k_\perp^2)]
      [m_N^2-m_{\omega N}^2(y,k_\perp^2)]}
      \nonumber\\
& & \left\{
	k^2\,
    \left[ g_{NN \rho}\,\left( -1 + y \right)  - 
      2\,f_{NN \rho}\,m_N \,y \right] \right\}
\eea
For pseudoscalar-vector interference terms, only helicity zero contributes
at leading twist.
\bea
f^0_{(\pi \rho) N/N}(y)&=&
\frac{3}{16\,m_\rho \, {\pi }^2\,{\left( -1 + y \right) }^3\,y^2}
\int_0^\infty 
\frac{d k_\perp^2 G_{\pi N}(y,k_\perp^2) G_{\rho N}(y,k_\perp^2)}
     {[m_N^2-m_{\pi N}^2(y,k_\perp^2)]
     [m_N^2-m_{\rho N}^2(y,k_\perp^2)]}
      \nonumber\\
& & \left\{
	g_{NN\pi}\,
    \left[ k^2 + {m_\rho}^2\,
       \left( -1 + y \right)  + m_N^2\,y^2
      \right] \right\} \nonumber \\
& & \left\{ - g_{NN\rho}\,
         m_N\,\left( -1 + y \right) \,y  +
       f_{NN\rho}\,\left[ - k^2\,
            \left( -2 + y \right)  + 
         m_N^2\,y^3 \right]  \right\} 
\eea
\bea
f^0_{(\pi \rho) \Delta/ N}(y)&=&
\frac{f_{N\Delta \pi} f_{N\Delta \rho} \, m_\rho}
{24\,m_\Delta \, {\pi }^2\,{\left( -1 + y \right) }^3\,y^2}
\int_0^\infty 
\frac{d k_\perp^2
		G_{\pi \Delta}(y,k_\perp^2)G_{\rho \Delta}(y,k_\perp^2)}
 {[m_N^2-m_{\pi \Delta}^2(y,k_\perp^2)]
  [m_N^2-m_{\rho \Delta}^2(y,k_\perp^2)]}
      \nonumber\\
& & \left\{
	k^4\,\left( 2 + y \right) \right. \nonumber \\
& & -2\,k^2\,\left[ 2\,m_\Delta\,m_N\,
          \left( -1 + y \right)  - 
         m_N^2\,{\left( -1 + y \right) }^2\,
          \left( 1 + y \right)  + 
         m_\Delta^2\,\left( -1 + 2\,y \right) 
         \right] \nonumber \\
& & \left.
	+\left[m_\Delta^2 - 
           m_N^2\,\left( -1 + y \right)^2
           \right]^2\,y
     \right\}
\eea
\bea
f^0_{(\pi^0 \omega) p/p}(y)&=&
\frac{- 3 \, g_{NN \pi} g_{NN \omega} \, m_N}
{16\,m_\omega \, {\pi }^2\,{\left( -1 + y \right) }^2\,y}
\int_0^\infty 
\frac{d k_\perp^2
		G_{\pi N}(y,k_\perp^2)G_{\rho N}(y,k_\perp^2)}
 {[m_N^2-m_{\pi N}^2(y,k_\perp^2)]
  [m_N^2-m_{\omega N}^2(y,k_\perp^2)]}
      \nonumber\\
& & \left[
k^2 + m_\omega^2\,
       \left( -1 + y \right)  + m_N^2\,y^2
       \right]
\eea

\newpage
\section*{Figure Captions}
\begin{description}
\item
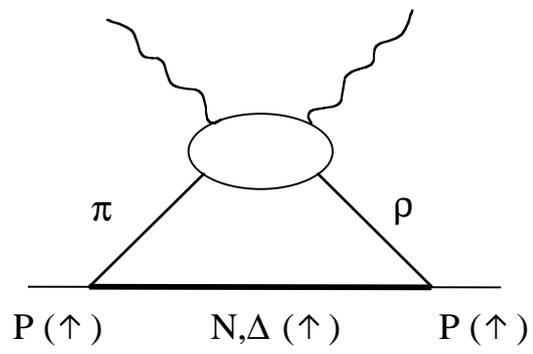
{Fig.~1.} 
Schematic illustration of interference contributions to the polarized
anti-quark distributions.
\item
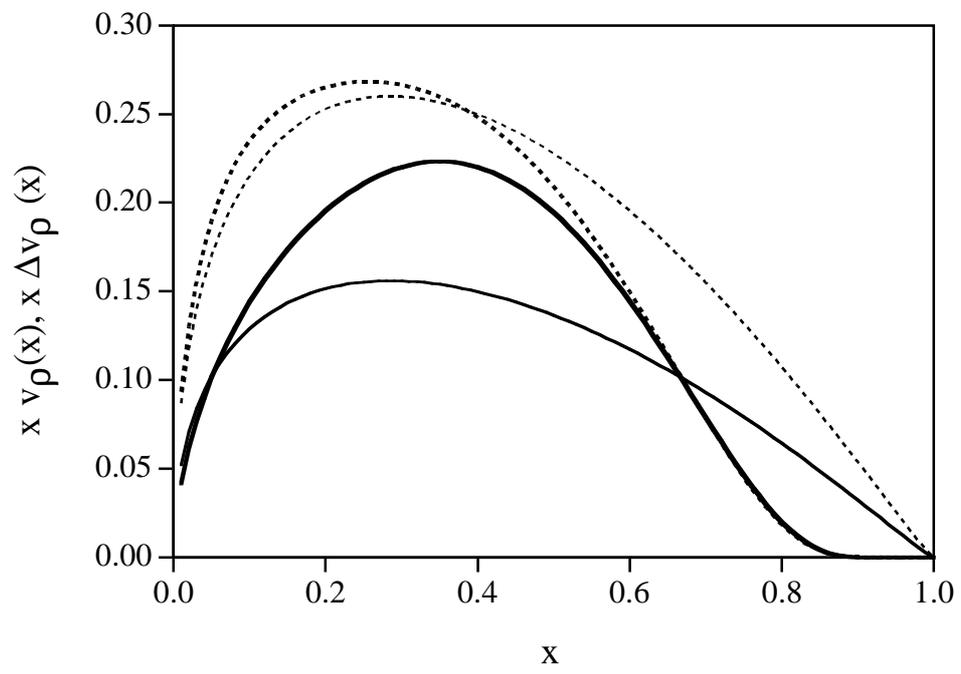
{Fig.~2.}
The polarized and unpolarized valence quark distributions of the $\rho$ meson 
at $Q^2=4$~GeV$^2$.
The thick dashed and solid curves are the bag model calculations for the unpolarized
and polarized distributions respectively.
The thin dashed curve is the unpolarized parton distribution
of the $\pi$ meson \cite{GRS99}.
The thick solid curve is the polarized parton distribution of the $\rho$ meson
using the ansatz $\Delta v_\rho(x)=0.6 \, v_\pi(x)$ \cite{RFriesS}.
\item
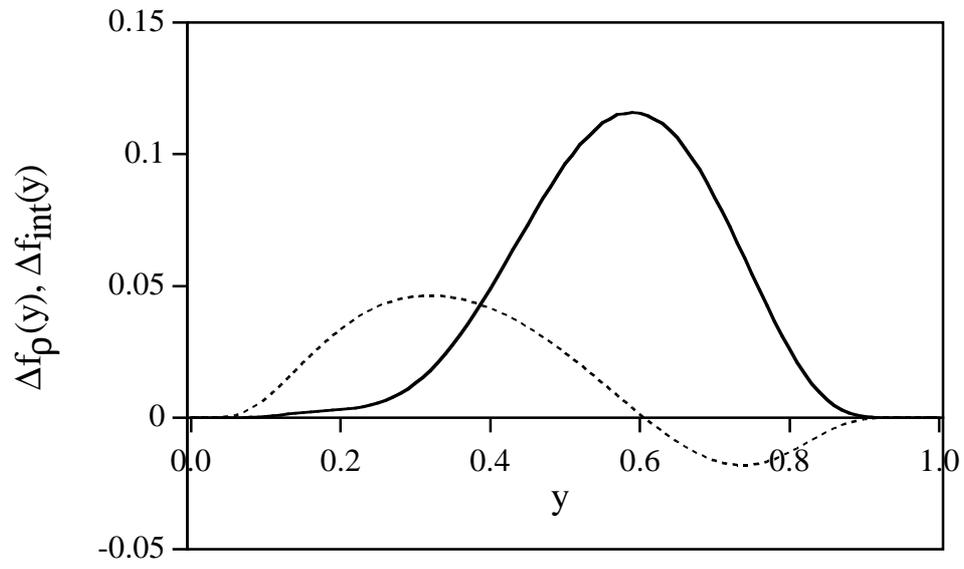
{Fig.~3.}
The polarized fluctuation functions for the vector meson (the solid curve)
and the interference terms (the dashed curve).
\item
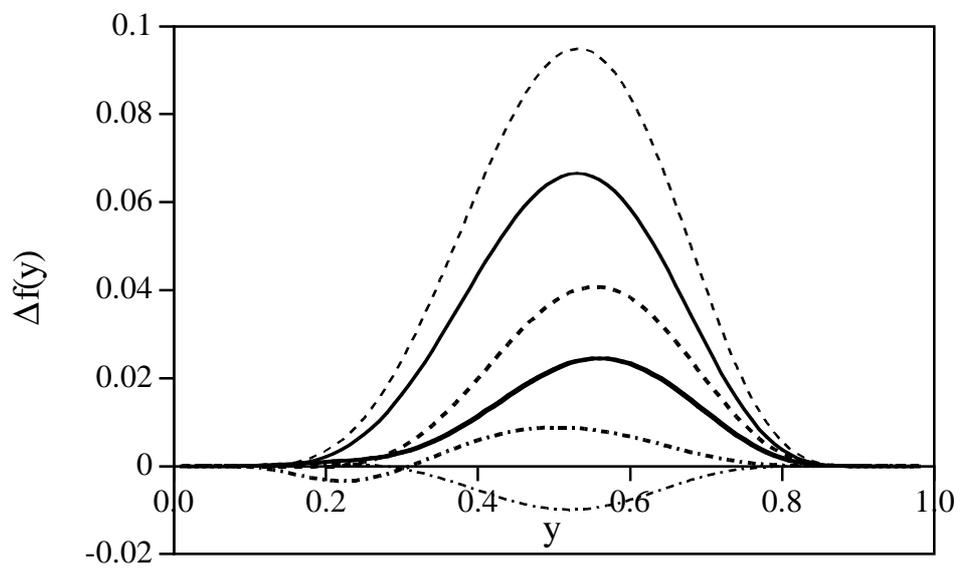
{Fig.~4.}
Comparision of polarized fluctuation functions, $\Delta f_{\rho NN}$ (dashed curves),
$\Delta f_{\rho \Delta N}$ (dashed-dotted curves), and
$\frac{2}{3}\Delta f_{\rho N N}-\frac{1}{3}\Delta f_{\rho \Delta N}$ (solid curves),
given in this work (thick curves) and in \cite{RFriesS} (thin curves).
\item
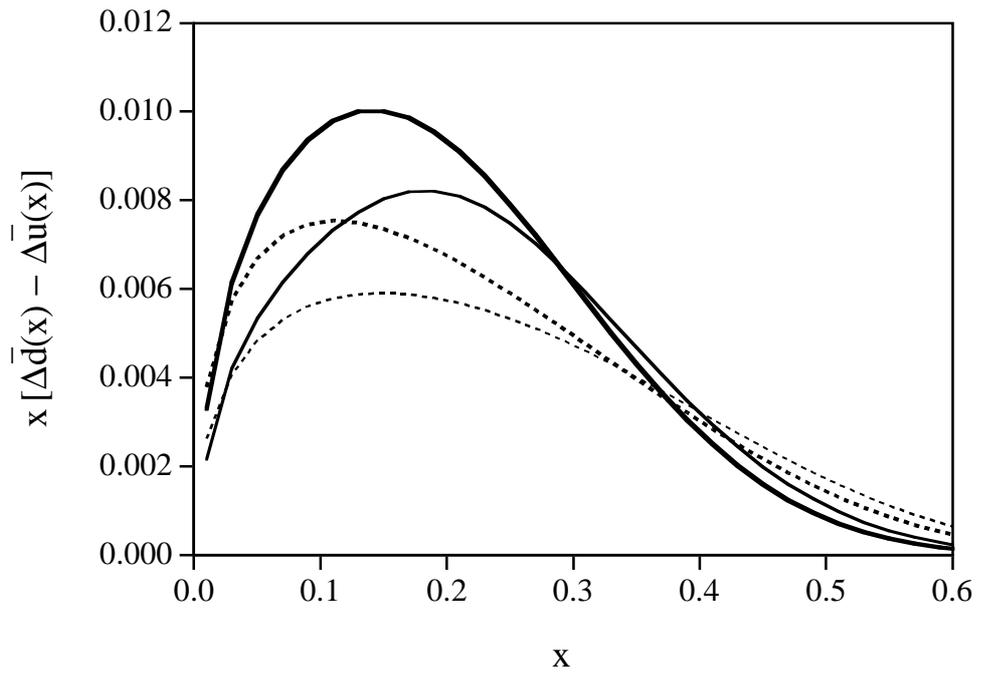
{Fig.~5.}
The flavour asymmetry of the anti-quark in the proton.
The solid curves are the predictions using $x\Delta v^{MIT}_\rho$, while
the dashed curves are obtained by using $ 0.6 \, xv_\pi(x)$.
The thin curves are the results without interference terms while the thick
curves are the results with interference terms.
\end{description}
%

\end{document}